\documentclass[sigconf]{acmart}
\usepackage{adjustbox}

\acmConference[QSIoT 2026]{Workshop on Quantum-Safe and Intelligent Security for IoT and Cyber-Physical Systems, co-located with IoT 2026}{November 17--20, 2026}{Newcastle upon Tyne, UK}
\acmYear{2026}
\copyrightyear{2026}
\setcopyright{none}
\renewcommand\footnotetextcopyrightpermission[1]{}
\begin{document}

\title{Testing TLS 1.3 PQC on constrained devices and microcontrollers}

\author{Olivier Gillot}
\affiliation{%
  \institution{Edinburgh Napier University}
  \city{Edinburgh}
  \country{United Kingdom}
}
\email{o.gillot@napier.ac.uk}

\author{William J Buchanan}
\affiliation{%
  \institution{Edinburgh Napier University}
  \city{Edinburgh}
  \country{United Kingdom}
}
\email{B.Buchanan@napier.ac.uk}

\author{Callum Turino}
\affiliation{%
  \institution{Edinburgh Napier University}
  \city{Edinburgh}
  \country{United Kingdom}
}
\email{C.Turino@napier.ac.uk}

\begin{abstract}
The UK, EU and US governments have set binding timelines for post-quantum cryptography migration of critical systems, with mandated deadlines ranging from 2028 to 2035. This paper specifically focuses on IoT-class devices and microcontrollers and the viability of using PQC to connect these devices in secure TLS 1.3 connections. Using wolfSSL, we benchmark a 36-combination TLS 1.3 algorithm matrix, pairing ML-KEM-512/768/1024 and X25519/ECC P-256 key exchange with ML-DSA-44/65/87, Ed25519, ECDSA, and RSA-PSS authentication, across four devices (Raspberry Pi Pico W \& Pico 2 W, ESP32, and a Raspberry Pi Zero), measuring both handshake latency (1,000 iterations per combination) and energy consumption. We found that in capable devices, all choices in the TLS matrix gave similar results, but that in IoT-class devices, PQC performance is highly implementation- and architecture-dependent: the common assumption that elliptic-curve key exchange must be cheaper than post-quantum lattice-based key exchange does not hold in this wolfSSL-based implementation, where the best full-PQC combination (ML-KEM-512 + ML-DSA-44) completes TLS handshakes 41\% faster than the equivalent classical-modern pairing (X25519 + RSA-PSS) on the Pico W (up to 2.12x as energy-efficient, depending on the paired authentication algorithm), and 8.9x faster on the ESP32. Full-PQC, ML-KEM-512/ML-DSA-44, is not only possible but energy-competitive with or superior to Hybrid or Elliptic Curve alternatives on the most constrained devices, though ML-DSA's variable signing time means applications with strict latency-predictability budgets must accept this variance as the cost of full quantum-safe authentication. We further show that the IETF's draft-specified X25519+ML-KEM-768 hybrid incurs approximately the combined cost of its two components in our measurements, rather than offering a performance compromise, making pure ML-KEM the more practical choice on hardware lacking Curve25519 assembly support.
\end{abstract}

\keywords{post-quantum cryptography, ML-KEM, ML-DSA, TLS 1.3, hybrid key exchange, IoT security, embedded systems, wolfSSL}

\ccsdesc[500]{Security and privacy~Cryptography}
\ccsdesc[500]{Security and privacy~Embedded systems security}
\ccsdesc[300]{Security and privacy~Key management}
\ccsdesc[300]{Networks~Network protocols}

\maketitle

\section{Introduction}

As Buchanan puts it, "Q-Day gets nearer" \cite{buchanan2026qday} -- keeping encryption secure is going to need new methods and new algorithms. For decades, RSA, and then Elliptic Curves have kept the internet secure, however, once a cryptographically relevant quantum computer (CRQC) exists, Shor's algorithm~\cite{shor1994algorithms} will allow it to break these methods of encryption. The current threat is simple: anything encrypted with quantum unsafe algorithms and recorded has the potential to be decrypted in the future. Hence, recording any encrypted data today means this data is not future-proof for prying eyes, a situation known as the "harvest-now, decrypt-later" (HNDL), which has critical implications for extra sensitive information. This is where protecting TLS 1.3 is critical. Once a CRQC exists, Shor's algorithm applied to the stored key-exchange values (X25519/ECC P-256 ECDH) recovers the private scalar, reconstructs the shared secret that the session used, and the stored ciphertext falls open. This is why TLS is the precise target of HNDL: TLS's key exchange is deliberately transmitted in the clear as part of the protocol, which is what makes it interceptable in the first place.

There has been an acknowledgment of the threat by governmental bodies. The UK has laid down a three-phase timetable for transition, with high-priority services expected to complete migration by 2031 \cite{ncsc2025migrating}, the EU has taken similar steps for 2035 \cite{ec2025pqc} and the US government has led the change by standardising a range of secure algorithms in 2024 \cite{csrc2024post} and a plan for a full transition by 2035 \cite{moody2024transition}.

In a previous paper \cite{gillot2026energy}, we focused our attention on generating keys and signatures to assess the energy efficiency of the newly standardised methods FIPS 203 (ML-KEM), FIPS 204 (ML-DSA) and FIPS 205 (SLH-DSA) on a range of constrained Linux machines. This paper goes further by directly assessing full TLS 1.3 exchanges for PQC viability on a Linux Raspberry Pi Zero as a comparison point and on microcontrollers: the Raspberry Pi Pico W \& Pico 2 W and the ESP32.

In our previous work, we had noticed that ML-KEM key generation and ML-DSA signing, the core algorithms for PQC TLS 1.3, performed well on even our most constrained machine, the Pi 1, with our OpenSSL 3.5 benchmark. However, in the IoT world, the chips used usually do not come with a full OS, and have considerably less clock speed and RAM to work with.

IoT-class devices matter because they equip many critical infrastructures in the medical, industrial and defence industries. They commonly are transmitting whatever data they collect via Wi-Fi or Bluetooth, and therefore these communications can be captured and recorded. It is trivial to install a listening device. Encrypting the data at the cloud edge is enticing, since servers are far more capable machines than embedded chips, but it is not sufficient to guarantee the safety of the data before it reaches the cloud.

Whilst OpenSSL is the standard library used for encrypting communications for servers, wolfSSL has been built from inception as a library to enable encryption on bare-metal machines and microcontrollers~\cite{wolfssl2026about} which explains why we selected it to build our custom benchmark.

TLS 1.3 \cite{rfc8446} has become the standard for secure communications, offering reduced latency and a significantly reduced attack surface compared to TLS 1.2. Leveraging wolfSSL's implementation of these algorithms, full PQC handshakes are now possible on machines configured with the appropriate algorithms.

In this work, we assess two main metrics: energy and speed. Both matter for practical reasons: if a method is too slow, it will not be viable for normal TLS exchange, and if a method needs too much energy, it won't be efficient on devices that are often low-powered battery operated. Selecting the right methods for TLS exchanges is critical to both the safety and the reliability of the communication.

\section{Related Work}

The work done here is in direct continuity of our previous paper \cite{gillot2026energy}. We built an OpenSSL benchmark to test energy and timings for key generation and signatures across classical, modern and PQC algorithms on a range of Raspberry Pis. The initial idea of testing energy consumption of PQC algorithms was found in Patterson~\cite{patterson2025energy}, who tested different algorithms on a Raspberry Pi 5. The second most important contribution to our idea was the work done by Turino~\cite{turino2025pqc}, whose idea to test TLS handshakes has been fundamental to this work.

Tasopoulos et al.~\cite{tasopoulos2023energy} demonstrated a full post-quantum TLS 1.3 exchange on a proper IoT device, an STM Nucleo Cortex-M4 board, using wolfSSL, including energy measurements. Kumar \cite{kumar2022post} provides a broad survey of the PQC landscape, and his implementation challenges section explicitly identifies resource-constrained devices as requiring "intense customisation and feasibility study" for PQC deployment. This work directly addresses that identified gap through empirical TLS 1.3 benchmarking across a 36-combination algorithm matrix on microcontrollers.

\begin{itemize}
  \item Schöffel et al. \cite{schoffel2021energy}: showed PQC KEMs are energy-feasible on a constrained Cortex-M4 without hardware accelerators; this work confirms and extends that finding to TLS handshakes across the Pico W, Pico 2 W, and ESP32.
  \item Roma et al. \cite{roma2021energy}: benchmarked NIST Round 2 PQC energy on a desktop Intel i7 using software power estimation; this work addresses their acknowledged gap by bringing physical measurements to genuinely low-end embedded hardware.
  \item Paquin et al. \cite{paquin2020benchmarking}: benchmarked pre-standardisation PQC in TLS under emulated network degradation on high-end cloud hardware; this work provides the complementary axis of constrained compute on capable networks.
  \item Sarıbaş and Tonyalı \cite{saribacs2022performance}, Bozhko et al. \cite{bozhko2023performance}: measured PQC TLS 1.3 packet overhead over Ethernet, BLE and WiFi; network performance is complementary to this work's focus on local compute and energy.
  \item Algar-Fernandez et al. \cite{algar2026benchmarking}: benchmarked standardised PQC timing on commodity Windows CPUs; this work addresses their explicitly acknowledged gap of microcontroller and ARM deployment environments.
\end{itemize}

Bürstinghaus-Steinbach et al. \cite{burstinghaus2020post} were, to their knowledge, the first to combine PQC, TLS, and embedded systems in a single evaluation, integrating pre-standardisation Kyber and SPHINCS+ into mbed TLS across a Raspberry Pi 3, an ESP32, an industrial field option card, and an LPC microcontroller. One finding is particularly relevant to this work: they note that the ESP32 provides hardware acceleration for big integer arithmetic, which benefits SECP256R1/ECDHE operations specifically, explaining the performance split this work also observes between ECC P-256 and X25519 on the ESP32. Their work was conducted on TLS 1.2, without energy measurements, and with pre-standardisation algorithm variants; this work extends theirs by testing the standardised FIPS 203/204 implementations over TLS 1.3, adding energy measurement, across a 36-combination algorithm matrix.

Anastasova et al. \cite{anastasova2024fully} present, to their knowledge, the first fully hybrid TLS 1.3 implementation in wolfSSL on an embedded device, a Cortex-M4 STM32F413, combining Curve448 and Crystals-Kyber1024 for key exchange. This makes it a direct methodological predecessor to this work: same library, same hybrid concept, same microcontroller class. However, their communication runs over UART rather than WiFi, and the work predates standardisation, using pre-FIPS Kyber and Dilithium names; this work extends this foundation to standardised FIPS 203/204 implementations, WiFi-connected devices, and energy measurement.

Pajkos et al. \cite{pajkos2025esp32} implemented a lightweight TLS 1.3 client on the ESP32-C6, a RISC-V-based variant of the ESP32 family, and report X25519 key pair generation at 14.33~ms. This study's own measurement of the same operation with wolfSSL on the Xtensa-based ESP32 is 379.65~ms (Table~\ref{tab:matrix-esp32}), over 26$\times$ slower; this independent comparison supports that the bottleneck this work identifies is ISA-specific rather than an ESP32-brand limitation.

Castiglione et al. \cite{castiglione2024integrating} implement Dilithium-5 (ML-DSA-87) on the ESP32 Xtensa LX6, and their signing times range from 165ms to 754ms, a nearly 5x spread across identical operations, directly corroborating the high coefficient of variation observed in ML-DSA combinations in this work's ESP32 TLS results.

Samwel et al. \cite{samwel2018breaking} demonstrate a differential power analysis attack that breaks Ed25519 in wolfSSL, the exact library used in this work, requiring only 4,000 power traces. Ed25519 appears as an authentication algorithm across six of the 36 TLS combinations benchmarked here, on hardware where physical access and power measurement is trivial; performance benchmarking alone is insufficient justification for choosing it in IoT deployments using wolfSSL.

Joseph et al. \cite{joseph2022transitioning} make the security case for hybrid cryptography as the recommended migration path, arguing that combining a classical and a post-quantum algorithm produces a system whose security is lower-bounded by the stronger of the two. This is the rationale behind X25519+ML-KEM-768, the hybrid key exchange tested in this work.

Taken together, this body of work establishes that PQC TLS on constrained hardware is feasible, but each study examines only a fragment of the picture: a single device, a single algorithm family, a single metric, or pre-standardisation algorithm variants. No prior work has evaluated the full FIPS 203/204 standardised algorithm set across a systematic TLS 1.3 matrix, on multiple microcontroller architectures simultaneously, measuring both handshake latency and energy consumption over a real WiFi connection. That is precisely the gap this work fills.

\section{Methodology}

Four devices were selected to cover a deliberate spectrum: from a 264~KB bare-metal microcontroller to a 32-bit Linux single-board computer, spanning three ISA families (ARM Cortex-M, Xtensa, and ARM11/ARMv6) and two runtime environments (bare-metal and Linux). This range allows direct comparison of algorithm performance across the constraints that matter most for IoT deployments: clock speed, available memory, ISA-specific library optimisations, and the presence or absence of an operating system.

The Raspberry Pi Pico~W and Pico~2~W are bare-metal ARM microcontrollers: wolfSSL runs directly on the hardware with no operating system. The Pico~W uses the RP2040 chip with a dual-core Cortex-M0+ at 125~MHz and only 264~KB of SRAM, the most constrained device in the test suite, and the one where memory limits directly determine which algorithms can even run. The Pico~2~W upgrades to the RP2350 with a dual-core Cortex-M33 at 150~MHz and 520~KB SRAM. The M33 core matters beyond the clock speed and memory increase: Cortex-M0+ implements the ARMv6-M architecture, which supports only six 32-bit Thumb-2 instructions rather than the full instruction set~\cite{arm2010armv6m}, whereas Cortex-M33 implements ARMv8-M Mainline with full Thumb-2 support~\cite{arm2020cortexm33}. wolfSSL's hand-written Curve25519 assembly ({\urlstyle{tt}\path{wolfcrypt/src/port/arm/thumb2-curve25519.S}}) is gated on full Thumb-2 availability, so it applies to Cortex-M33 but not Cortex-M0+, making the Pico~2~W substantially better-integrated with the library and directly explaining several performance gaps observed in the results.

The ESP32 DevKitC~V1 is one of the most widely deployed IoT microcontrollers in the world, and was included specifically because it uses a non-ARM architecture, the Xtensa~LX6, at a higher clock speed (240~MHz) than either Pico. This makes it an ideal test for ISA-specific behaviour: wolfSSL ships assembly-optimised paths for ARM Cortex-M but has no equivalent for Xtensa, meaning the ESP32 falls back to generic~C for Curve25519-family operations despite its clock advantage. The ESP32 also includes hardware acceleration for AES, SHA-2, and RSA arithmetic, but not for Curve25519, a distinction that becomes central to interpreting the results. It runs FreeRTOS via the Arduino framework.

The Raspberry Pi Zero WH is the Linux baseline: a Broadcom BCM2835 SoC with a single-core ARM1176JZF-S at 1~GHz (ARM11, ARMv6, 32-bit) and 512~MB of RAM, running Raspberry Pi OS (Linux, Bookworm). It is the only device in the study with a full operating system, a standard TCP/IP networking stack, and virtual memory, none of which the three bare-metal microcontrollers have, and its role is to isolate which of this paper's findings are properties of bare-metal, memory-constrained execution specifically, and which persist even once those constraints are removed.

wolfSSL is the standard cryptographic library for embedded devices and microcontrollers, and was the favoured option for this work as it allows a single codebase to be deployed across all devices tested. Custom-built in C, it is designed to run directly on hardware without an operating system, making it particularly well-suited for bare-metal PQC deployment on devices such as the Pico and ESP32. wolfSSL has integrated ML-KEM and ML-DSA natively and deprecated the liboqs library.

We use ``full-PQC'' throughout to mean that both the TLS key-exchange and authentication primitives use standardised PQC algorithms (ML-KEM and ML-DSA); this does not imply that every element of the surrounding PKI, including the certificate chain itself, is post-quantum.

\begin{table*}[t]
\caption{The 36-combination TLS 1.3 algorithm matrix. Rows: key exchange (traditional, hybrid, full-PQC). Columns: authentication signature. Cell notation drops the \texttt{ML-} prefix for brevity (\texttt{KEM} = ML-KEM, \texttt{DSA} = ML-DSA). Hybrid row = X25519+ML-KEM-768.}
\label{tab:tls-matrix}
\centering
\setlength{\tabcolsep}{2pt}
\begin{tabular}{lcccccc}
\toprule
Key Exchange & RSA-PSS 2048 & ECDSA (P-256) & Ed25519 & ML-DSA-44 & ML-DSA-65 & ML-DSA-87 \\
\midrule
ECDHE (P-256) & ECDHE/RSA    & ECDHE/ECDSA    & ECDHE/Ed25519    & ECDHE/DSA-44    & ECDHE/DSA-65    & ECDHE/DSA-87    \\
X25519        & X25519/RSA   & X25519/ECDSA   & X25519/Ed25519   & X25519/DSA-44   & X25519/DSA-65   & X25519/DSA-87   \\
ML-KEM-512    & KEM-512/RSA  & KEM-512/ECDSA  & KEM-512/Ed25519  & KEM-512/DSA-44  & KEM-512/DSA-65  & KEM-512/DSA-87  \\
ML-KEM-768    & KEM-768/RSA  & KEM-768/ECDSA  & KEM-768/Ed25519  & KEM-768/DSA-44  & KEM-768/DSA-65  & KEM-768/DSA-87  \\
ML-KEM-1024   & KEM-1024/RSA & KEM-1024/ECDSA & KEM-1024/Ed25519 & KEM-1024/DSA-44 & KEM-1024/DSA-65 & KEM-1024/DSA-87 \\
Hybrid        & Hybrid/RSA   & Hybrid/ECDSA   & Hybrid/Ed25519   & Hybrid/DSA-44   & Hybrid/DSA-65   & Hybrid/DSA-87   \\
\bottomrule
\end{tabular}
\end{table*}

The benchmark shares a common codebase, compiled into three platform-specific versions: Linux 32-bit, Pico firmware, and ESP32 firmware. The difference lies in data collection: the Pi Zero is capable of writing results directly to a \texttt{.csv} file, which was then transferred to the analysis machine. The Picos and ESP32 cannot write files to persistent storage, so a screen recording script captured the terminal output during each test run; a Python parser then extracted the relevant data and produced a \texttt{.csv} file for analysis.

The FNIRSI FNB58 was selected to collect the energy measurements, via its convenient Wh counter. Energy measurements collected by the FNB58 are saved in a proprietary \texttt{.cfn} format; a conversion script transforms these into \texttt{.csv} for analysis. It was connected between the device under test and a Windows machine running the companion software (FNIRSI-UsbMeter, v0.0.7), the only platform the proprietary FNB58 software supports. The USB-meter program was configured to record energy (Wh counter) at 50 samples per second, in 180-second runs, for every method tested (key generation, encapsulation/decapsulation, signing/verifying) and for every key exchange $\times$ signature combination in the TLS matrix. Each combination is measured in its own dedicated 180-second run, not apportioned from a shared or mixed-operation measurement: only that one combination executes during the run, repeating back-to-back for the full 180 seconds. The FNB58 records total energy consumed (Wh) over that run, and the benchmark separately records the number of iterations completed in the same window; iterations per Joule is then the iteration count divided by the run's total energy (J = Wh $\times$ 3,600). This reports the average energy cost of one handshake for that specific combination, inclusive of whatever idle time falls between consecutive handshakes within the run (e.g. TCP connection setup and teardown), rather than an oscilloscope-isolated measurement of the cryptographic operation alone; it therefore reflects the realistic amortised cost of running that combination under sustained back-to-back load, which is what a persistent deployment would actually draw. Before operational measurements, each machine received three baseline runs to establish idle energy consumption.

To verify the validity of the TLS exchanges, TCPDump was used to capture TLS-related network traffic and Wireshark to analyse the captured packets, confirming the cryptographic suite negotiated in each handshake.

Energy figures were calculated on two bases: total energy, the operational run's raw energy consumption with no adjustment, and delta energy, that run's energy minus the mean of the three idle-baseline runs for that device, isolating the energy cost attributable to the cryptographic operation itself. Total and delta energy are strongly correlated across every device tested (Pearson correlation coefficient $r$ between 0.78 and 1.00, computed within each operation category), so the relative ranking of algorithms on a given device is the same under either basis.

Timing readings were necessary to validate the energy readings and provide a fuller picture of algorithm performance, including standard deviation, median values (the most reliable measure for this test scenario), 95th and 99th percentiles, and minimum and maximum values. 1,000 iterations were used for every operation on every machine.

Every one of the iterations per combination is a completely fresh handshake, as if the client had never connected before: none of TLS 1.3's shortcuts for reconnecting to a known server (session resumption, session tickets, 0-RTT early data) are used, so every measurement pays the full cryptographic cost rather than a cheaper reconnect cost. Iterations run one at a time, never in parallel, over a normal home WiFi network between two separate physical machines: the device under test, and a single server, a Lenovo ThinkPad P14s running Ubuntu, that answered every combination in the study. We time only the handshake itself: the clock starts the instant \texttt{wolfSSL\_connect()} is called and stops the instant it returns successfully, so the reported latency excludes the TCP connection setup beforehand and the teardown afterward.

One setting needs explaining directly, since it is easy to misread as cutting a corner: certificate chain-of-trust checking is switched off (\texttt{WOLFSSL\_VERIFY\_NONE}). This only removes the bookkeeping step of confirming a certificate was signed by an already-trusted authority, work that would otherwise require provisioning a correct trust store on every device for all 36 algorithm combinations. It does not remove any cryptography: the operation actually under measurement, the client cryptographically verifying the server's \texttt{CertificateVerify} signature, still executes in full on every handshake. Each of the six authentication algorithms also has its own dedicated certificate and key, generated specifically for that algorithm, so no measurement mixes one algorithm's certificate with another's result.

The source code for the benchmarks, the recorded energy files, the pcap verification files and the full results, for both timings and energy, are freely available at \url{https://github.com/gillo7/qsiot2026-pqc-tls-benchmark}.

\section{Results}

\subsection{Pico W: ML-KEM-512 vs Classical Key Exchange}

\begin{figure*}[t]
\centering
\includegraphics[width=0.8\linewidth]{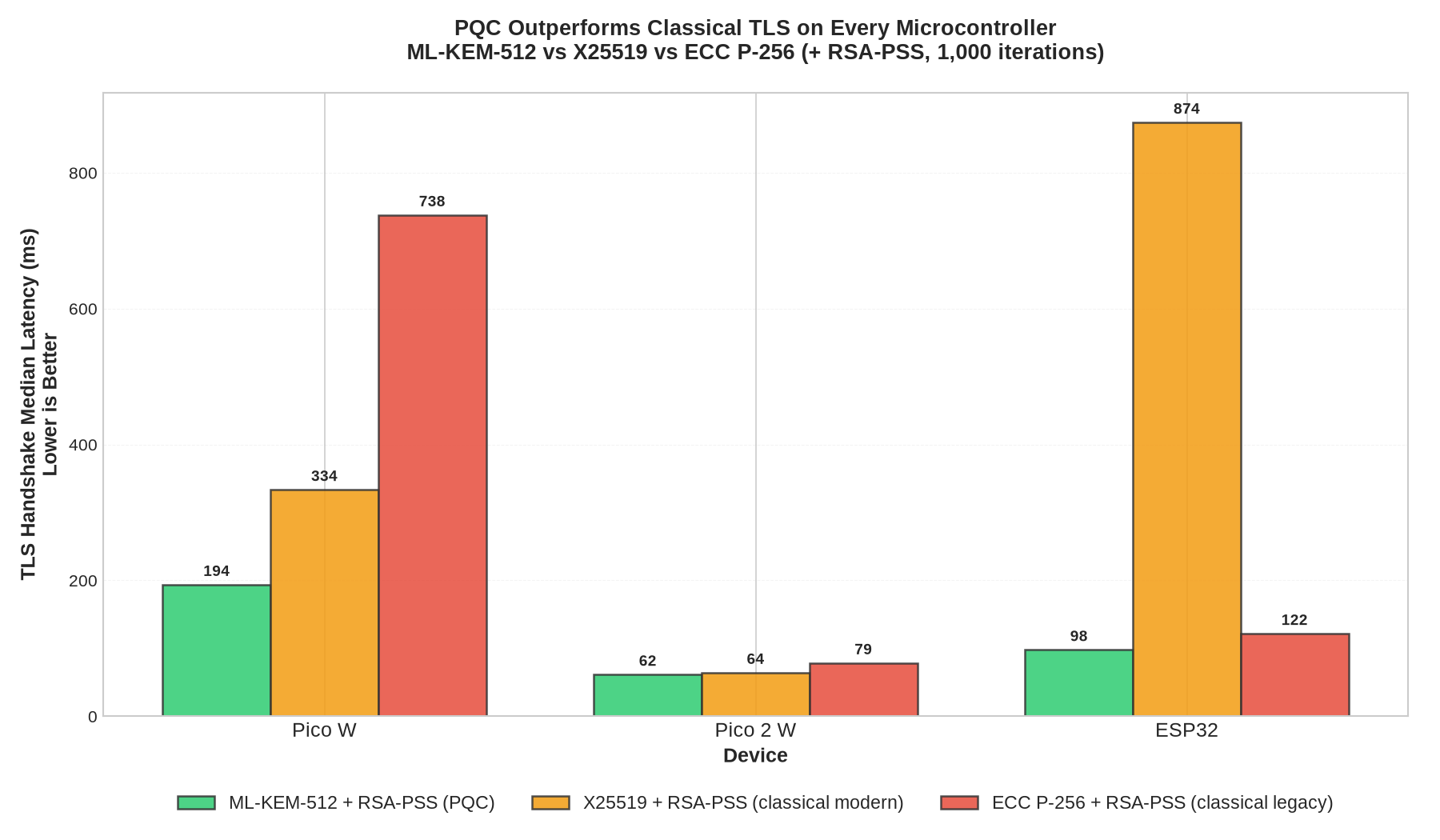}
\Description{Bar chart comparing TLS handshake latency for PQC and classical key-exchange combinations across the three microcontrollers, showing PQC combinations consistently faster.}
\caption{PQC vs classical TLS latency, all microcontrollers.}
\label{fig:pqc-vs-classical}
\end{figure*}

The headline finding is that ML-KEM-512 is faster than both ECDHE and X25519 for TLS on every microcontroller tested, across every paired authentication algorithm (Figure~\ref{fig:pqc-vs-classical}). The best full-PQC combination (ML-KEM-512 + ML-DSA-44 at 197.44 ms) is 41\% faster than the equivalent classical-modern pairing (X25519 + RSA-PSS at 333.69 ms) and 3.7x faster than the legacy classical pairing (ECC P-256 + RSA-PSS at 737.97 ms) (Table~\ref{tab:matrix-picow}). The energy advantage is sharper on some pairings than others: against ECDSA (P-256) authentication specifically, ML-KEM-512 reaches 154.0 iterations/joule versus X25519's 72.7, 2.12x the efficiency for the same signature algorithm. This advantage does not automatically extend to larger ML-KEM parameter sets: on Pico 2 W, X25519 + RSA-PSS (64.5 ms, 356.8 iterations/joule) outperforms both ML-KEM-768 + RSA-PSS (69.2 ms, 327.2 iterations/joule) and ML-KEM-1024 + RSA-PSS (82.9 ms, 275.3 iterations/joule) (Table~\ref{tab:matrix-pico2w}). The result is device- and parameter-specific rather than a blanket rule: ML-KEM-512 is the consistent winner, but the assumption that any post-quantum key exchange is automatically cheaper than classical key exchange does not hold once larger security levels are selected.

\begin{table*}[t]
\centering
\caption{Full TLS 1.3 matrix -- Pico W: median latency (ms) / efficiency (iterations per joule, delta energy). Key exchange: ECDHE = ECDHE (P-256), KEM-512/768/1024 = ML-KEM-512/768/1024, Hybrid = X25519+ML-KEM-768. Authentication: RSA = RSA-PSS 2048, ECDSA = ECDSA (P-256), DSA-44/65/87 = ML-DSA-44/65/87. OOM = out of memory; the Pico W is the only device in this study where any combination fails this way -- Pico 2 W, ESP32, and Pi Zero (Tables 3--5) complete every combination.}
\label{tab:matrix-picow}
\begin{tabular*}{\textwidth}{l@{\extracolsep{\fill}}cccccc}
\toprule
Key Exchange & RSA & ECDSA & Ed25519 & DSA-44 & DSA-65 & DSA-87 \\
\midrule
ECDHE & 738.0/91.6 & 757.9/48.9 & 753.0/82.8 & 742.7/37.5 & OOM & OOM \\
X25519 & 333.7/102.3 & 349.9/72.7 & 346.5/98.7 & 336.4/105.0 & OOM & OOM \\
KEM-512 & 193.9/130.8 & 208.7/154.0 & 206.9/153.4 & 197.4/141.9 & OOM & OOM \\
KEM-768 & 209.3/144.7 & 224.5/125.7 & 220.8/141.8 & OOM & OOM & OOM \\
KEM-1024 & 235.8/134.0 & 252.1/135.3 & 248.4/128.6 & OOM & OOM & OOM \\
Hybrid & 473.1/87.4 & 498.7/78.2 & 485.6/95.2 & OOM & OOM & OOM \\
\bottomrule
\end{tabular*}
\end{table*}

\begin{table*}[t]
\centering
\caption{Full TLS 1.3 matrix -- Pico 2 W: median latency (ms) / efficiency (iterations per joule, delta energy). Abbreviations as Table~\ref{tab:matrix-picow}.}
\label{tab:matrix-pico2w}
\begin{tabular*}{\textwidth}{l@{\extracolsep{\fill}}cccccc}
\toprule
Key Exchange & RSA & ECDSA & Ed25519 & DSA-44 & DSA-65 & DSA-87 \\
\midrule
ECDHE & 78.6/308.5 & 85.4/183.8 & 83.1/274.7 & 110.2/180.9 & 129.2/205.2 & 159.9/144.1 \\
X25519 & 64.5/356.8 & 72.2/293.7 & 68.7/340.1 & 96.2/237.7 & 115.1/205.9 & 146.1/126.5 \\
KEM-512 & 61.7/374.4 & 68.4/258.9 & 65.2/362.0 & 93.3/204.7 & 112.3/152.5 & 142.6/134.0 \\
KEM-768 & 69.2/327.2 & 75.7/213.0 & 72.0/330.3 & 99.8/217.4 & 118.8/227.3 & 150.2/151.5 \\
KEM-1024 & 82.9/275.3 & 90.3/265.1 & 86.1/278.1 & 113.2/227.1 & 133.1/200.6 & 162.8/142.5 \\
Hybrid & 91.3/277.6 & 98.1/259.9 & 95.0/270.8 & 122.2/220.6 & 141.6/159.6 & 172.3/99.2 \\
\bottomrule
\end{tabular*}
\end{table*}

\begin{table*}[t]
\centering
\caption{Full TLS 1.3 matrix -- ESP32: median latency (ms) / efficiency (iterations per joule, delta energy). Abbreviations as Table~\ref{tab:matrix-picow}.}
\label{tab:matrix-esp32}
\begin{tabular*}{\textwidth}{l@{\extracolsep{\fill}}cccccc}
\toprule
Key Exchange & RSA & ECDSA & Ed25519 & DSA-44 & DSA-65 & DSA-87 \\
\midrule
ECDHE & 121.8/61.1 & 126.3/52.1 & 970.7/22.3 & 133.4/51.1 & 156.4/45.8 & 185.6/40.5 \\
X25519 & 874.2/18.4 & 999.4/16.2 & 1761.1/12.1 & 1007.3/15.2 & 1149.7/14.6 & 1124.1/14.1 \\
KEM-512 & 98.1/71.6 & 106.8/59.5 & 954.5/18.6 & 115.2/56.5 & 145.7/48.2 & 172.0/39.4 \\
KEM-768 & 104.6/72.7 & 114.7/57.9 & 964.1/19.3 & 121.2/56.3 & 146.6/48.4 & 175.8/40.6 \\
KEM-1024 & 124.8/57.7 & 131.5/51.1 & 975.2/19.7 & 137.1/54.1 & 165.1/47.0 & 192.8/43.7 \\
Hybrid & 997.1/20.7 & 1052.7/17.8 & 1744.6/11.5 & 1026.5/15.6 & 1051.0/15.7 & 1040.1/15.4 \\
\bottomrule
\end{tabular*}
\end{table*}

\begin{table*}[t]
\centering
\caption{Full TLS 1.3 matrix -- Pi Zero: median latency (ms) / efficiency (iterations per joule, delta energy). Abbreviations as Table~\ref{tab:matrix-picow}.}
\label{tab:matrix-pizero}
\begin{tabular*}{\textwidth}{l@{\extracolsep{\fill}}cccccc}
\toprule
Key Exchange & RSA & ECDSA & Ed25519 & DSA-44 & DSA-65 & DSA-87 \\
\midrule
ECDHE & 21.5/248.1 & 23.2/246.0 & 20.2/263.7 & 24.5/235.3 & 27.8/210.6 & 33.4/172.1 \\
X25519 & 18.3/242.4 & 19.7/254.6 & 16.3/292.0 & 20.7/243.5 & 24.1/224.9 & 29.5/195.1 \\
KEM-512 & 16.9/302.4 & 18.3/293.2 & 14.3/327.3 & 19.8/265.9 & 22.5/241.6 & 28.6/207.2 \\
KEM-768 & 18.8/277.1 & 20.2/270.2 & 17.5/298.4 & 22.2/255.4 & 26.2/220.8 & 31.2/196.3 \\
KEM-1024 & 20.7/242.9 & 23.7/245.0 & 19.5/267.6 & 23.4/233.0 & 27.1/210.4 & 32.5/179.5 \\
Hybrid & 22.9/226.0 & 24.7/227.8 & 21.5/234.5 & 26.0/217.2 & 29.3/195.6 & 35.6/168.0 \\
\bottomrule
\end{tabular*}
\end{table*}

\subsection{ESP32: Library Integration vs Clock Speed}
\label{sec:isa}

The dominant finding on the ESP32 is the collapse of X25519 performance. In the TLS matrix, combinations using neither X25519 nor Ed25519 (excluding the X25519 and Hybrid key-exchange rows, and the Ed25519 authentication column) achieve between 39.4 and 72.7 iterations per joule, while X25519-based combinations, pure or hybrid, cap at 20.7 iterations per joule (Hybrid/RSA), 1.9 to 3.5 times less. This is not a hardware limitation but an implementation gap. wolfSSL's Curve25519 optimisations target ARM Cortex-M assembly; on the Xtensa LX6, no equivalent path exists and the implementation falls through to a generic C fallback regardless of clock speed. ML-KEM's lattice arithmetic requires no ISA-specific assembly to achieve competitive performance and scales proportionally across all architectures tested.

\begin{figure*}[t]
\centering
\includegraphics[width=0.8\linewidth]{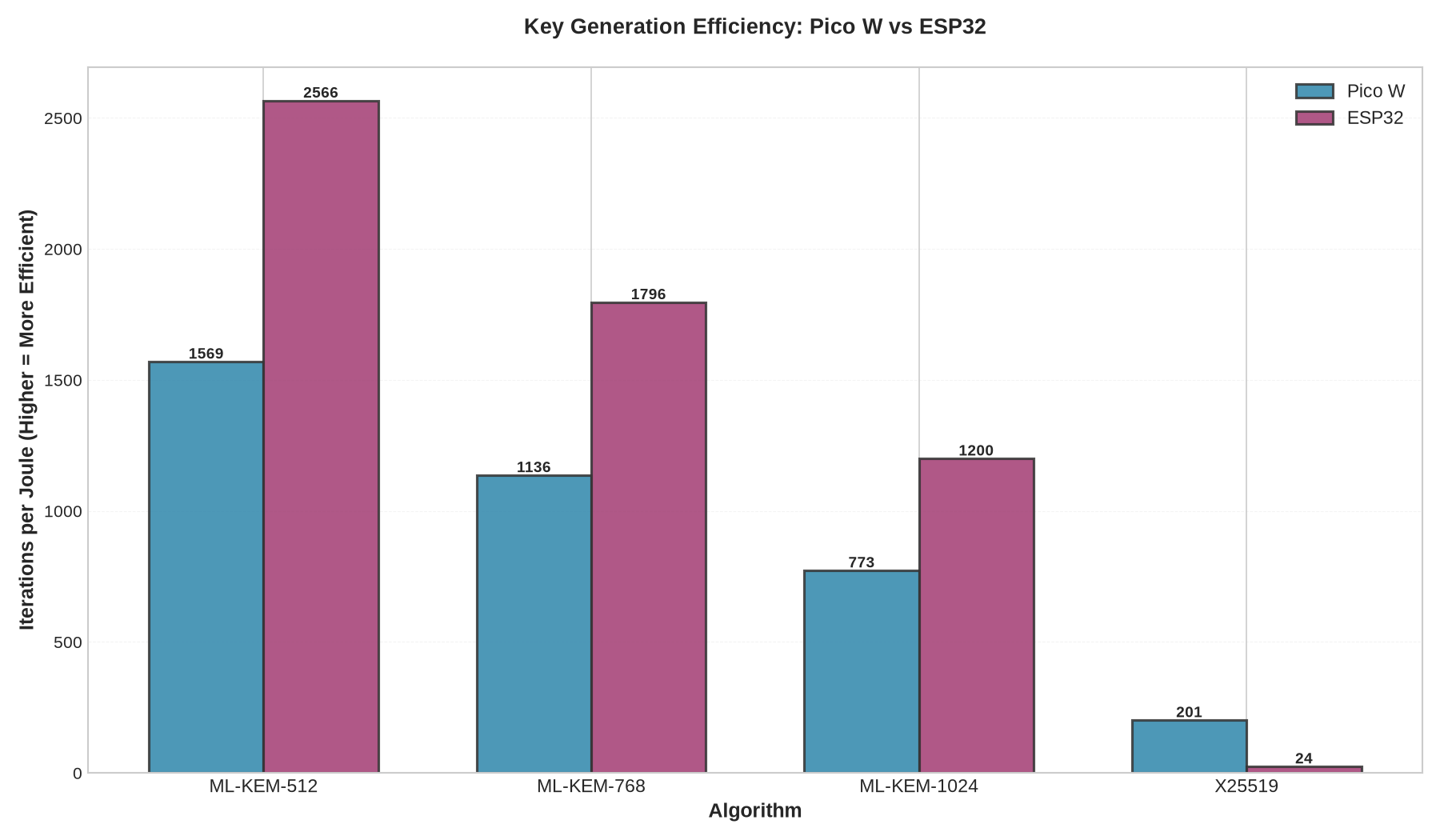}
\Description{Bar chart comparing key-generation energy efficiency between the Pico W and ESP32 across ML-KEM and X25519, showing ML-KEM far ahead on both devices while X25519 collapses specifically on the ESP32.}
\caption{Key generation efficiency, Pico~W vs ESP32.}
\label{fig:keygen-isa}
\end{figure*}

\subsection{The Hybrid Combiner: Additive Cost, Partial Protection}

Kwiatkowski and Langley's IETF draft on hybrid key agreement for TLS 1.3~\cite{kwiatkowski2024ecdhemlkem} defines X25519 + ML-KEM-768 as one of its named hybrid groups. In our measurements, the computational cost of this hybrid implementation was approximately the combined cost of its X25519 and ML-KEM-768 components: on Pico W, the hybrid ranges from 473~ms (+ RSA-PSS) to 499~ms (+ ECC P-256), 2.4--2.6x worse than pure ML-KEM-512 for the same authentication step. On ESP32, the hybrid ranges from 997~ms (+ RSA-PSS) to 1,745~ms (+ Ed25519), adding the X25519 penalty on top of the ML-KEM-768 overhead. Pico 2 W fares better at 91--172~ms across hybrid combinations, given its lower X25519 overhead. Until wolfSSL ships Xtensa and Cortex-M0+ assembly for Curve25519, pure ML-KEM remains the more practical choice for latency-constrained microcontrollers.

The second limitation is one of scope rather than cost: the hybrid construction protects only the key exchange, not the authentication step. The \texttt{CertificateVerify} signature is generated with whatever signature algorithm is separately negotiated (RSA-PSS, ECDSA, or Ed25519 in the classical case), and none of those are made quantum-safe merely by pairing them with a hybrid key exchange. A quantum adversary who cannot break $ss_{\text{hybrid}}$ to read the traffic can, given a CRQC, still forge the server's classical signature and impersonate the server outright, since the certificate's cryptographic binding to its private key relies on exactly the discrete-logarithm or factorisation hardness that Shor's algorithm removes. This is the harvest-now-attack-the-certificate-later problem: confidentiality of the session is protected by the hybrid, but authenticity of the endpoint is not, unless the signature algorithm is also upgraded to ML-DSA or another quantum-safe algorithm.

\subsection{Memory as a Hard Constraint}

The Pico W's most immediate constraint is memory, and Table~\ref{tab:matrix-picow} shows it is driven by the combined footprint of key exchange and signature rather than either alone: the smaller key exchanges (ECDHE, X25519, ML-KEM-512) complete successfully through ML-DSA-44 and only exhaust the 264~KB SRAM at ML-DSA-65/87, while the larger ML-KEM-768, ML-KEM-1024, and Hybrid combinations already run out of memory at ML-DSA-44. The Pico~W is the only device in this study to run out of memory on any combination in the matrix; the Pico~2~W, ESP32, and Pi~Zero each complete every TLS combination tested, including those requiring ML-DSA-87 signing.

\section{Discussion}

Library integration is the performance-determining factor, not raw clock speed. The single most consistent predictor of algorithm performance across devices is not MHz but whether wolfSSL has assembly-optimised routines for the target ISA. Pico 2 W completes X25519/Ed25519/Hybrid combinations across the matrix faster than the faster-clocked ESP32 because Cortex-M33 has the assembly; Xtensa does not. This is a practical finding for IoT deployments: selecting hardware for PQC should include assessment of cryptographic library ISA support, not just clock speed and SRAM. ML-KEM exhibited more consistent cross-architecture performance in our wolfSSL configurations, whereas Curve25519 performance was strongly affected by the availability of ISA-specific optimisation. ML-KEM's lattice operations scale with clock speed across ARM, Xtensa, and x86-equivalent Linux SBCs with consistent relative performance in this benchmark. For deployments targeting multiple hardware platforms, ML-KEM's portability is a significant operational advantage.

Our results demonstrate that ML-KEM-based TLS 1.3 is computationally feasible on constrained IoT hardware under the tested configurations, with the wolfSSL library. On every device tested, ML-KEM-512 TLS completes faster than the classical X25519 standard it is designed to replace, not as a future promise, but as a measurable result across 1,000-iteration latency runs and 180-second energy runs, on hardware ranging from a low-cost microcontroller to a Linux single-board computer. ML-KEM-512 also generally matched or exceeded X25519's energy efficiency, though the magnitude and even the direction of that advantage depends on the paired authentication algorithm and the device: on Pico~2~W, X25519 is more energy-efficient than ML-KEM-512 for three of the six authentication pairings (Table~\ref{tab:matrix-pico2w}). These results complicate the conventional assumption that elliptic-curve key exchange must be cheaper than post-quantum lattice-based key exchange on constrained hardware: that assumption held for the larger ML-KEM-768 and ML-KEM-1024 parameter sets on some devices (Section 4.1), but not for ML-KEM-512, the parameter set most relevant to lightweight IoT deployment. The primary implementation risk is not algorithmic but environmental: algorithm-library-hardware co-selection is critical, and the ISA support gap in wolfSSL for Xtensa makes the ESP32 a poor host for any Curve25519-family algorithm. Computational feasibility is not production readiness, however: this work does not establish interoperability, certificate lifecycle management, secure boot, firmware update mechanisms, side-channel and fault-injection resistance, long-term key storage, or behaviour under degraded network conditions or concurrent application workload, all of which still stand between this result and a deployed device. What this work does establish is that compute and energy cost are not the barrier they might be assumed to be, and that the remaining engineering effort belongs in matching the chosen library's assembly support to the deployment target's instruction set.

\section{Limitations and Future Work}

These benchmarks measure a specific implementation (wolfSSL) on specific hardware under controlled WiFi conditions. Results for X25519 and Ed25519 on ESP32 are specific to the absence of Xtensa assembly in the wolfSSL build used; a vendor-optimised implementation could change the picture for Curve25519 on that platform.

This paper also does not address whether the extra bytes PQC key exchange and signatures add on the wire could offset its compute-side advantage: ML-DSA and larger ML-KEM parameter sets are meaningfully larger than their classical equivalents, and this work only measures handshake latency end-to-end, not compute and communication cost separately. We have packet-capture and timing data on larger Raspberry Pi models from the wider project (Section~3): on a Pi~3, the full-PQC combination (ML-KEM-512 + ML-DSA-44) sends 3.34$\times$ more bytes than the classical-modern pairing (X25519 + RSA-PSS), yet its median handshake time was not slower (13.49~ms vs.\ 14.00~ms). This is interesting, but we have not yet repeated it on this paper's own four devices, the Pico~W, Pico~2~W, ESP32, and Pi~Zero, so whether communication cost stays negligible relative to compute on these microcontrollers specifically is left as an open question for future work.

Nor could this work test FALCON, HQC, or FrodoKEM: none are currently implemented in wolfSSL, which has also deprecated the liboqs library that would otherwise have provided a path to test them~\cite{wolfssl2025liboqs}.

The memory figures used throughout this paper are each device's nominal total SRAM, not measured free heap. The benchmark did not instrument free-heap logging on any device, so the usable heap remaining at the point of the Pico~W's out-of-memory failures, and how much of its 264~KB SRAM the networking stack and wolfSSL itself reserve before the application allocates its own key material, could not be reported. Logging free heap immediately before each handshake would sharpen the memory-constraint claims in Section~4.4.

Finally, this study has used a FNIRSI FNB58, a capable consumer-grade USB power meter, but one that lacks the reliability of a state-of-the-art oscilloscope, which could more efficiently isolate individual operations and provide more accurate power metrics.

\section{Conclusion}

The answer to whether post-quantum cryptography works on constrained IoT devices is yes, but not as a blanket algorithmic law: PQC performance on constrained IoT hardware is highly implementation- and architecture-dependent, and the conventional assumption that elliptic-curve key exchange must be cheaper than post-quantum lattice-based key exchange does not hold under the realistic embedded configuration tested here. The core result is this: every microcontroller in the study completes ML-KEM-512 TLS 1.3 handshakes faster than the currently deployed X25519-based equivalents, from the 264 KB Pico W to a 32-bit Pi Zero, and ML-KEM-512 generally matched or exceeded X25519 in energy efficiency too, though on Pico 2 W the direction depends on the paired authentication algorithm. The margin is substantial (41\% lower latency on Pico W, 8.9x lower latency on ESP32), and traces to a specific, identifiable cause rather than a general algorithmic advantage: wolfSSL's X25519 implementation relies on ARM Cortex-M assembly that does not exist for Xtensa, and is absent for Cortex-M0+ at useful clock speeds. ML-KEM's lattice arithmetic, by contrast, requires no ISA-specific assembly paths to achieve competitive performance. This advantage is specific to ML-KEM-512: on Pico 2 W, X25519 outperforms the larger ML-KEM-768 and ML-KEM-1024 parameter sets in several pairings (Section 4.1), so the relative cost of PQC versus classical cryptography on constrained hardware is a property of the specific library, architecture, and parameter set, not the algorithm family in isolation. The cross-architecture consistency observed here reflects wolfSSL's optimisation coverage rather than an inherent algorithmic guarantee, but it is a real, measurable difference in this benchmark, and one an implementer can act on today.

\bibliographystyle{ACM-Reference-Format}
\bibliography{references}

\end{document}